\documentclass{epl}
\usepackage{amssymb}
\usepackage{psfig}
\usepackage{amsmath}
\usepackage{amsfonts}
\usepackage{graphicx}

\begin{document}

\title{Vacuum induced torque between corrugated metallic plates}
\author{Robson B. Rodrigues \inst{1} \and Paulo A. Maia Neto \inst{1} 
\thanks{E-mail: pamn@if.ufrj.br} \and A. Lambrecht\inst{2} \and S. Reynaud\inst{2} 
\thanks{E-mail: reynaud@spectro.jussieu.fr} }
\institute{ \inst{1} Instituto de F\'{\i}sica, UFRJ, Caixa Postal 68528, 
21941-972 Rio de Janeiro RJ, Brazil\\
\inst{2} Laboratoire Kastler Brossel,
CNRS, ENS, UPMC case 74, Campus Jussieu, 75252 Paris, France}
\shortauthor{R. B. Rodrigues \etal} 
\shorttitle{Vacuum induced torque between corrugated plates} 


\maketitle

\begin{abstract}
We study the torque arising between two corrugated metallic plates
due to the interaction with electromagnetic vacuum. 
This Casimir torque can be measured with torsion pendulum techniques for
separation distances as large as 1$\mu$m. It allows one to probe the
nontrivial geometry dependence of the Casimir energy in a configuration
which can be evaluated theoretically with accuracy. In the optimal
experimental configuration, the commonly used proximity force approximation 
turns out to overestimate the torque by a factor 2 or larger.
\end{abstract}

The Casimir effect \cite{Casimir} plays a major role in micro- and
nano-electromechanical systems (MEMS and NEMS) \cite{capasso}. Besides the
normal Casimir force between metallic or dielectric plates \cite{normal},
the observation of the lateral Casimir force between corrugated plates 
\cite{chen} opens novel possibilities of micro-mechanical control. 
The lateral force results from breaking the translational symmetry along 
directions parallel to the plates by imprinting periodic corrugations 
on both metallic plates. As rotational symmetry is also broken by this 
geometry, a Casimir torque should also arise when the corrugations are not 
aligned. In the present letter, we study this effect which provides one 
with a new mechanism of micro-mechanical control to be exploited in the 
design of MEMS and NEMS.

From the point of view of fundamental physics, this effect makes possible 
an accurate investigation of the non-trivial geometry dependence of
the Casimir energy. Tests of this geometry dependence performed up to date
have often been limited by the use of the proximity-force approximation (PFA)
in the calculations \cite{PFA}. Within this approximation, the energy is 
evaluated by simply averaging the Casimir energy between two planes 
over the distribution of local distances, which is certainly insufficient 
except for the limiting case of large corrugation wavelengths \cite{EPL2003}. 
Other calculations \cite{Emig} have gone farther than PFA while using a 
model of perfect reflection, which limited their validity to experiments 
performed with metallic mirrors at distances larger than a few $\mu$m.

Here we use the scattering approach \cite{Genet2003,scattering} extended 
to treat non-planar surfaces \cite{PRA2005,PRL2006}. This perturbative
approach is restricted by the assumption that corrugation amplitudes 
are smaller than  other length scales but it allows one to
go beyond the range of validity of the PFA, with arbitrary values
of the corrugation wavelengths with respect to plate separation and 
plasma wavelength of the metal. As shown below, it accurately describes  
configurations which are experimentally testable 
and where the Casimir energy has a non-trivial geometry dependence. 
This new effect induced by vacuum fluctuations may be small because
the corrugation amplitudes are small, but the torque detection is 
particularly interesting from this point of view, since it allows one
to use the exquisite sensitivity of torsion balances \cite{torsionbalances}.

The idea of using torsion techniques has already been proposed to measure
Casimir torques between anisotropic dielectric plates 
\cite{Parsegian,Barash,vanEnk,Torres}.
Recently, an experiment has been proposed to measure the torque with a
birefringent disk on top of a barium titanate plate \cite{Iannuzzi}.
According to the authors, the proposed measurement is feasible at a 
plate separation $L$ of 100nm. We show below that the torque between
corrugated metallic plates is up to three orders of magnitude larger than
the torque between anisotropic dielectric plates, for comparable values of
the separation distance and plate area. With realistic values for the
corrugation amplitudes $a_1$ and $a_2$ and wavelength $\lambda_C$, we
conclude that the torque measurement may be done at a distance $L$ of the
order of 1$\mu$m. This minimizes the spurious effects due to the
non-parallelism of the plates while making the contribution of interband
transitions negligible. We thus model the finite conductivity of the
metallic plates by the plasma model with a plasma wavelength $\lambda_P$.
We calculate the modification of the Casimir energy up to second
order in the corrugation amplitudes assumed to be smaller than other
length scales 
\begin{equation}  \label{perturb}
a_1, a_2 \ll L\,,\,\lambda_C\,,\, \lambda_P
\end{equation}  
This condition, allowing us to use perturbation theory, will be satisfied by the
numerical examples discussed below.

The surface profile functions for the two plates are denoted as $h_j(\mathbf{%
r})$ with $j=1,2$ labeling the two plates. They define the local heights,
counted as positive when corresponding to separation decreases with respect
to reference planes at $z=0$ and $z=L$. Both $h_j$ have zero spatial
averages, so that $L$ represents the average separation distance between the
two surfaces, and $\mathbf{r}=(x,y)$ collects the two transverse coordinates
defining the position on the plates.
The corrugated surfaces are described by non-specular reflection
coefficients that couple different field polarizations and momenta \cite%
{foots}. 
The second-order correction is then given by \cite{PRL2006} 
\begin{equation}  \label{energy}
\delta E_{\mathrm{PP}}=\int \frac{d^{2}\mathbf{K}}{(2\pi )^{2}} \mathcal{G}(%
\mathbf{K})H_{1}(\mathbf{K})H_{2}(-\mathbf{K}),
\end{equation}
where $H_j(\mathbf{K})$ is the Fourier transform of $h_j(\mathbf{r})$. The
response function $\mathcal{G}(\mathbf{K})$ does not depend on the direction
of the corrugation wavevector $\mathbf{K.}$ It is given in \cite{PRL2006} as
a function of the specular and non-specular reflection coefficients. Only
the crossed terms of the form $H_1H_2$ are kept in eq.~(\ref{energy}),
since terms quadratic in $H_1$ and in $H_2$ do not depend on the relative
angle between the plates, and hence do not contribute to the torque.

We assume the corrugations to have sinusoidal shapes $h_j(\mathbf{r}) =
a_j \cos( \mathbf{k}_j \cdot \mathbf{r} - k b_j)$ with corrugation
wavevectors $\mathbf{k}_j$ having the same modulus $k=2\pi/\lambda_C$. The
angle $\theta$ between $\mathbf{k}_1$ and $\mathbf{k}_2$ represents the
angular mismatch between the two corrugations and $\mathbf{k}_2$ is
supposed, for convenience, to be aligned along the direction of the $x-$axis.
The parameters $b_j$ represent lateral displacements with respect to the
configuration with a line of maximum height at the origin. With these
conventions, eq.~(\ref{energy}) yields 
\begin{equation}  \label{energy2}
\delta E_{\mathrm{PP}}=a_1\, \mathcal{G}(k) \,\mathrm{Re}\left[e^{i k
b_1}H_2(\mathbf{k}_1)\right]
\end{equation}
The integral yielding $H_2(\mathbf{k}_1)$ may be specified by considering that the corrugation $h_2$ is
restricted to a rectangular section of area $L_x L_y$ centered at $x=b_2,y=0.
$ We assume that both $L_x$ and $L_y$ are much larger than $L,$ so that
diffraction at the borders of the plates is negligible. The energy
correction per unit area is then given by 
\begin{eqnarray}  \label{energy3}
\frac{\delta E_{\mathrm{PP}}}{L_x L_y}&=&\frac{a_1a_2}{2}\, \mathcal{G}(k)
\cos\left(k b\right) \mathrm{sinc}(k L_y\sin\theta/2) \sum_{\epsilon=\pm} 
\mathrm{sinc}\left[k L_x(1+\epsilon \cos\theta)/2\right] \\
b&=& b_2\cos\theta-b_1 \quad,\quad \mathrm{sinc}(x)\equiv \frac{\sin(x)}{x}
\end{eqnarray}
$b$ is the relative lateral displacement along the direction of $\mathbf{k}_1
$; as expected by symmetry, the energy does not depend on displacements
perpendicular to $\mathbf{k}_1$. Eq.~(\ref{energy3}) satisfies reflection
symmetry around the $x$ and $y$ directions (invariance under the
transformations $\theta\rightarrow -\theta$ and $\theta\rightarrow \pi
-\theta$), which results from the fact that the corrugation lines have no
orientation. It contains as a special case ($\theta=0$)
the result for pure lateral displacement derived in Ref.~\cite{PRL2006}.

The energy variation with $b$ and $\theta$ given by eq.~(\ref{energy3}) is
universal, since the separation distance $L$ and additional parameters
characterizing the metallic surfaces determine only the global pre-factor $%
\mathcal{G}(k).$ In the limit of long corrugation lines, $k L_y\gg 1,$ we
may take the approximation $\theta \ll 1$ in eq.~(\ref{energy3}), since $%
\delta E_{\mathrm{PP}}$ is negligible otherwise. If $L_x$ is smaller or of
the order of $L_y,$ we find from eq.~(\ref{energy3}) 
\begin{equation}  \label{energy4}
\frac{\delta E_{\mathrm{PP}}}{L_x L_y}=\frac{a_1a_2}{2}\, \mathcal{G}(k)
\cos\left(k b\right) \, \mathrm{sinc}(k L_y\theta/2).
\end{equation}
Hence, the scale of variation of  $\theta$ is set by the parameter $1/(k
L_y)=\lambda_C/(2\pi\,L_y).$

\begin{figure}[ptb]
\centerline{\psfig{figure=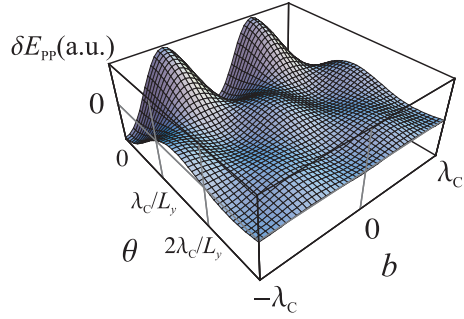,width=6cm}}
\caption{ Variation of Casimir energy (in arbitrary units) with the rotation
angle $\protect\theta$ and the lateral displacement $b.$}
\label{plot3D}
\end{figure}

In Fig.~\ref{plot3D}, we plot $\delta E_{\mathrm{PP}}$  (in arbitrary units)
as a function of $b$ and $\theta.$ Since $\mathcal{G}(k)$ is negative \cite%
{PRL2006}, the Casimir energy is minimum at $\theta=0$ and $%
b=0,\lambda_C,2\lambda_C,...,$ corresponding to the geometry with aligned
corrugations and the local separation distance having the maximum
variation amplitude. There are also shallow wells around $\theta \approx
1.43\lambda_C/L_y$ (minimum of $\mathrm{sinc}(k L_y \theta/2)$)  and $%
b=\lambda_C/2,3\lambda_C/2,...$ If we start from $\theta=b=0$ and rotate
plate 2 around its center, we follow the line $b=0$ in Fig.~\ref{plot3D}.
This figure shows that for small angles the plate is attracted back to $%
\theta=b=0$ without sliding laterally. On the other hand, if the plate is
released after a rotation of $\theta > \lambda_C/L_y$ its subsequent motion
will be a combination of rotation and lateral displacement. In fact, the
energy correction vanishes at $\theta=\lambda_C/L_y$, defining the range of
stability of the configuration $b=\theta=0.$ Rotation is favored over
lateral displacements for $\theta <\lambda_C/L_y.$

\begin{figure}[ptb]
\centerline{\psfig{figure=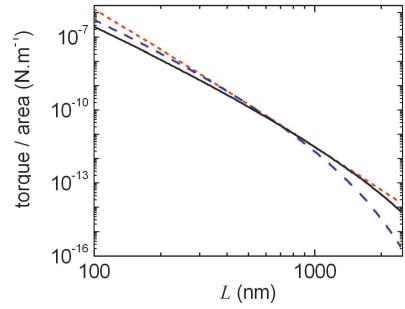,width=6cm}}
\caption{Maximum torque per unit area as a function of mean separation
distance. Corrugation amplitudes: $a_1a_2=200\,\mathrm{nm}^2,$ line length: $%
L_y=24\,\protect\mu\mathrm{m},$ plasma wavelength: $\protect\lambda_P=137\,%
\mathrm{nm}.$ Solid line: $\protect\lambda_C=2.4\,\protect\mu\mathrm{m};$
dashed line: $\protect\lambda_C=1.2\,\protect\mu\mathrm{m};$ dotted line: $%
\protect\lambda_C=2\protect\pi L/2.6$ (optimum value). }
\label{res}
\end{figure}

We now proceed to an evaluation of the torque which is given by 
\begin{equation}
\tau =-\frac{\partial }{\partial \theta }\,\delta E_{\mathrm{PP}}.
\end{equation}
It is maximum at $\theta= 0.66\lambda_C/L_y$ where it is given by 
\begin{equation}  \label{torque}
\frac{\tau}{L_xL_y}= 0.109\, a_1a_2\, k \mathcal{G}(k)\,L_y.
\end{equation}
As could be expected, this maximal torque per unit area is proportional to
the length $L_y$ of the corrugation lines, which provides the scale for the
moment arm. 
In order to plot the value of the torque (\ref{torque}) on Fig.~\ref{res}, 
we take the plasma wavelength corresponding to gold for both plates ($\lambda_P=137$nm). 
We chose corrugation amplitudes $a_1=a_2=14\,\mathrm{nm}$ obeying condition 
(\ref{perturb}) and leading to $a_1a_2=200\,\mathrm{nm}^2$ 
(to be compared with $a_1a_2= 472\, \mathrm{nm}^2$ in the lateral force 
experiment \cite{chen}, where $a_1$ and $a_2$ were unequal) and $L_y= 24 \,\mu$m. 
Different values for $a_1a_2$ and $L_y$ can also be derived from these results 
through a mere multiplication by suitable pre-factors (see eq.~\ref{torque}), 
provided that these values satisfy $a_1,a_2\ll L, \lambda_C$ and $L_y\gg \lambda_C/(2\pi).$
The first condition is at the heart of our perturbative approach, whereas 
the second condition can be relaxed by going
back to the more general result (\ref{energy3}).

The dashed line in Fig.~\ref{res} corresponds to the corrugation period 
$\lambda_C=1.2\,\mu\mathrm{m}$ of the lateral force experiment \cite{chen}.
At $L=100\,\mathrm{nm},$ we find in this case  $\tau/(L_xL_y)=5.2\times
10^{-7}\,\mathrm{N.m^{-1}}$, approximately three orders of magnitude
larger than the torque per unit area for anisotropic plates calculated in
Ref.~\cite{Iannuzzi} for the most favorable configuration at the same 
separation distance. 
The much larger figures found in our case should certainly allow one to 
perform the experiment at larger separation distances. 
For $L\, \overset{>}{\scriptscriptstyle\sim}\, 1\,\mu\mathrm{m},$
Fig.~\ref{res} shows that the torque for $\lambda_C=1.2\,\mu\mathrm{m}$
starts to decrease exponentially. However, by selecting longer corrugation
periods, one also finds measurable orders of magnitude in this range of
distances.

At any given value of $L$, the torque between corrugated plates can be made
larger by chosing the corrugation period so as to maximize $k \mathcal{G}(k).$ 
In the range of separation distances shown on Fig.~\ref{res}, this corresponds
to $k\approx 2.6/L$ or $\lambda_C\approx 2\pi L/2.6$. The torque is thus given 
by the dotted line, which provides upper bounds for its magnitude.
At $L=1\,\mu\mathrm{m},$ the optimum value is $\lambda_C=2.4 \,\mu\mathrm{m},
$ corresponding to the solid line in Fig.~\ref{res}. In this case, we find $%
\tau/(L_xL_y)=3.0 \times 10^{-12}\,\mathrm{N.m^{-1}}.$ As for the relevant
angle scales, the stability threshold is at $\theta=\lambda_C/L_y = 0.1 \,%
\mathrm{rad} = 5.7^{\mathrm{o}}$ and the maximum torque at $\theta=3.8^{%
\mathrm{o}}.$ If rotation is to be observed from the position of the tip of
the rectangular plate, then the relevant parameter is the arc described by
the tip $(L_y/2) \theta \sim \lambda_C,$ which is independent of $L_y.$

\begin{figure}[ptb]
\centerline{\psfig{figure=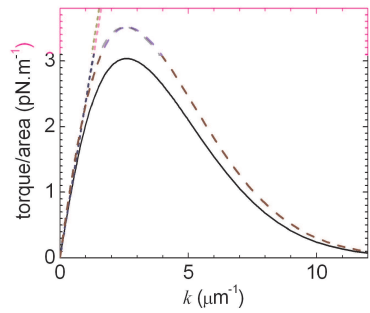,width=6cm}}
\caption{ Maximum torque per unit area as a function of $k=2\protect\pi/%
\protect\lambda_C$ with $L=1\,\protect\mu\mathrm{m}.$ Additional parameters
are chosen as in Fig.~2. The solid line corresponds to the theory presented
in this letter. We also plot the results for perfect reflectors (dashed
line) and PFA with plasma model (dotted line). }
\label{tvsk}
\end{figure}

Under such optimum conditions, the torque probes a non-trivial geometry
dependence of the Casimir energy, since the PFA thus grossly overestimates
the effect, as shown now. The PFA holds for smooth surfaces corresponding to
the limit $\lambda_C\rightarrow \infty.$ It is recovered from eq.~(\ref%
{torque}) at the limit $k\rightarrow 0$, where the response function
satisfies the general condition \cite{PRL2006} $\mathcal{G}(0)=e_{\mathrm{PP}%
}^{\prime\prime}(L),$ where $e_{\mathrm{PP}}$ is the Casimir energy per unit
area for parallel planes. In this limit, the effect of geometry is trivial
because the PFA directly connects the non-planar geometry to the more
commonly studied configuration of parallel planes. We thus find a result
determined by the Casimir energy for this simpler geometry: 
\begin{equation}  \label{torquePFA}
\left(\frac{\tau}{L_xL_y}\right)_{\mathrm{PFA}}= 0.109\, a_1a_2\, k e_{%
\mathrm{PP}}^{\prime\prime}(L)\,L_y.
\end{equation}
In order to compare with the PFA, we plot our results for the torque as a
function of $k$ for $L=1\,\mu\mathrm{m}$ in Fig.~\ref{tvsk} (solid line), 
and with all other parameters as in Fig.~\ref{res}. As discussed above, the
torque is maximum at $k = 2.6/L = 2.6\,\mu\mathrm{m}^{-1}.$ We also show the
values obtained from the model with perfect reflectors (dashed line). They
overestimate the torque by $16\%$ near the peak region.

\begin{figure}[ptb]
\centerline{\psfig{figure=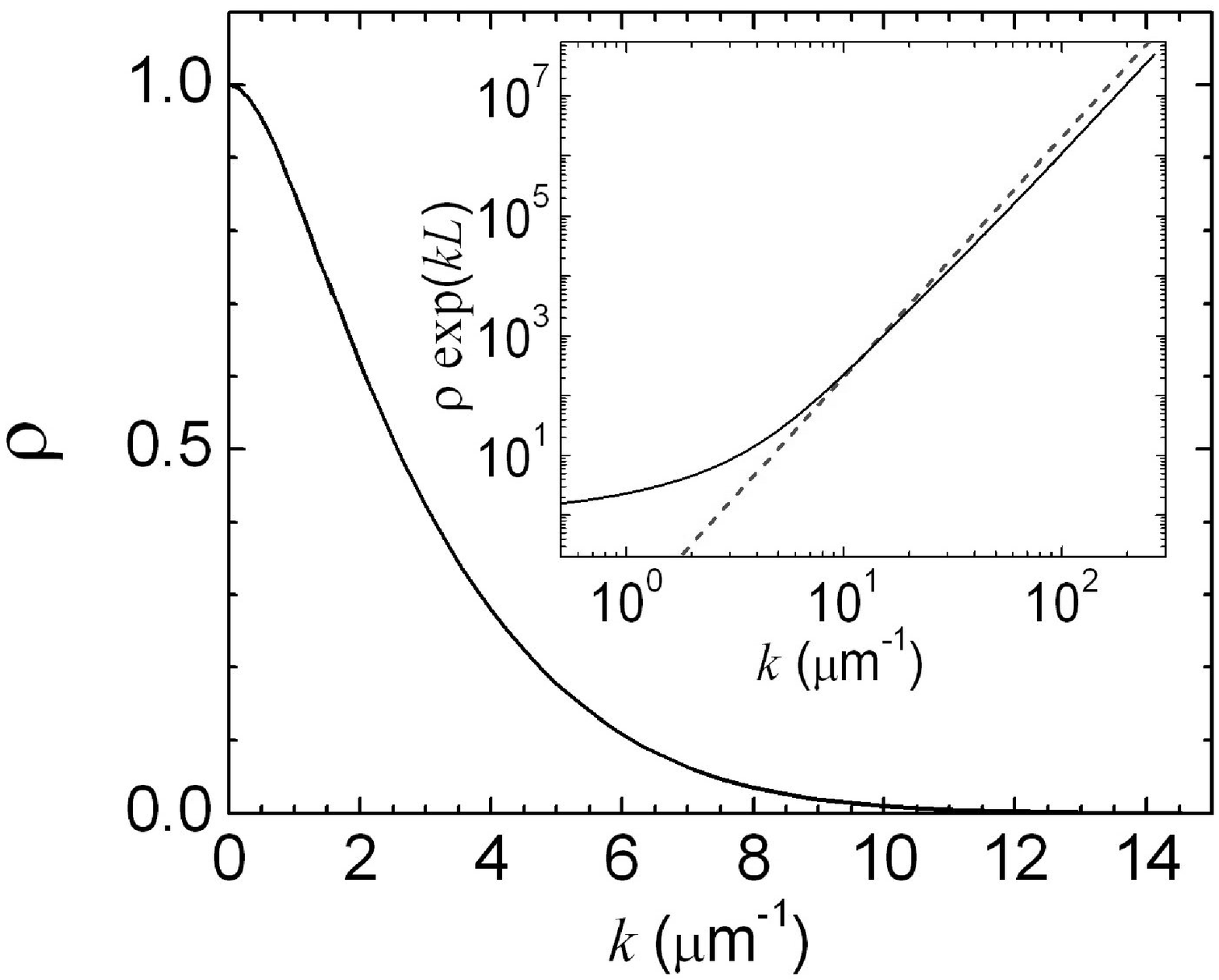,width=7cm}}
\caption{ Variation of $\protect\rho$ versus $k$. Parameters are chosen as
in Fig.~3. Inset: variation of $\protect\rho(k)\,\exp(kL)$ (solid line) and
perfectly-reflecting high-$k$ limit (dashed line). }
\label{rho-gamma}
\end{figure}

According to eq.~(\ref{torquePFA}), the torque grows linearly with $k$ in
the PFA (dotted line). It is thus worth increasing the value of $k$ and
hence going out the region of validity of PFA. At the peak value $k = 2.6/L
= 2.6\,\mu\mathrm{m}^{-1},$ the PFA overestimates the torque by $103\%.$ As
the torque decays for larger values of $k,$ the PFA rapidly becomes less and
less accurate. The discrepancy is measured by the ratio 
\begin{equation}
\rho(k) = \frac{\mathcal{G}(k)}{e_{\mathrm{PP}}^{\prime\prime}(L)}=\frac{%
\mathcal{G}(k)}{\mathcal{G}(0)}
\end{equation}
At $k=0,$ the exact and PFA values coincide ($\rho(0)=1$). For $k>0,$ $%
\rho(k)<1,$ as in the numerical example of Fig.~\ref{rho-gamma}. Thus, the
PFA always overestimates the torque and the discrepancy increases with $k,$
as expected, since smaller values of $k$ correspond to smoother surfaces.

For $k\gg 1/L,$ $\mathcal{G}(k)$ and $\rho(k)$ decay exponentially to zero.
Such behavior has a simple interpretation within the scattering approach 
\cite{PRL2006}. The reflection by the corrugated surfaces produce lateral
wavevector components of the order of $k.$ After replacing the integral over
real frequencies by an integral over the imaginary axis in the complex
plane, we obtain propagation factors, representing one-way propagation
between the plates, of the order of $\exp(-kL).$

In order to analyze the large-$k$ behavior in more detail, we plot $%
\rho(k)\, \exp(kL)$ in the inset of Fig.~\ref{rho-gamma}. In the
intermediate range $1/L \ll k \ll 2\pi/\lambda_P = 46\,\mu\mathrm{m}^{-1},$ $%
\rho(k)$ approaches the high-$k$ limit of the model with perfect reflectors 
\cite{Emig}, drawn as the dashed line in the inset of Fig.~\ref{rho-gamma}, 
\begin{equation}
\rho(k) = \frac{2}{\pi^4}\,(kL)^4 \,\exp(-kL)\,\,\,\,\,\;\;\;\mbox{(perfect
reflectors)}
\end{equation}
As $k$ approaches $2\pi/\lambda_P,$ $\rho(k)$ stays more and more below the
result for perfect reflectors, reaching the following limit for very large
values of $k$ (not shown on the figure), 
\begin{equation}
\rho(k) = \frac{5}{2}\,kL \,\exp(-kL)\,\,\,\,\,\;\;\;(1/L\ll 2\pi/\lambda_P
\ll k)
\end{equation}
Thus, the perfect-reflecting regime is reached only when the plasma
wavelength is the shortest length scale in the problem (apart from the
corrugation amplitudes). Just taking the limit $L\gg \lambda_P$ is not
sufficient for recovering the limit of perfect reflectors \cite{Emig}. This
has again a very simple interpretation in the scattering approach. When $%
L\gg \lambda_P,$ the relevant input field modes are perfectly reflected
because they have wavevectors of order of $1/L.$ However, diffraction by the
corrugated surfaces produces wavevectors of the order of $k,$ which are
poorly reflected by the plates if $k \gg 1/\lambda_P.$

In conclusion, we have studied the torque induced by vacuum fluctuations
between corrugated metallic plates. This Casimir torque may provide one 
with a new mechanism of micro-mechanical control to be exploited in the 
design of MEMS and NEMS. It could as well be a new observable of interest 
to test the geometry dependence of the Casimir energy. 
We recover the PFA from our theory in the limit $L\ll \lambda_C,$ 
but deviations become rapidly important as $L/\lambda_C$ is increased. 

The torque is up to three orders of magnitude larger than the torque 
between anisotropic dielectric plates for comparable distance and area. 
This should allow for an experimental observation of the Casimir torque 
at separation distances around $L = 1 \,\mu$m, using corrugation periods 
of the same order of magnitude ($\lambda_C= 2.4 L$). 
In this configuration, the PFA grossly overestimates the torque
by a factor of the order of 2. Moreover, it predicts an algebraic decay of
the torque as $L$ is increased past the optimum value, whereas the exact
decay is actually exponential. This experiment would provide, for the first
time, direct evidence of the nontrivial geometry dependence of the Casimir
effect.

\acknowledgments

R.B.R. and P.A.M.N. thank O. N. Mesquita for discussions and FAPERJ, CNPq
and Institutos do Mil\^enio de Informa\c c\~ao Qu\^antica e Nanoci\^encias
for financial support. A.L. acknowledges partial financial support by the European Contract STRP
12142 NANOCASE.


\begin{thebibliography}{99}
\bibitem{Casimir} CASIMIR H. B. G., \textit{Proc. K. Ned. Akad. Wet.}, 
\textbf{51}  (1948) 793.

\bibitem{capasso} CHAN H. B., AKSYUK V. A., KLEIMAN R. N., BISHOP D. J. and
CAPASSO, F., \textit{Science}, \textbf{291} (2001) 1941.

\bibitem{normal} KLIMCHITSKAYA G. L. \textit{et al}, \textit{J. Phys. A:
Math Gen.} \textbf{39} (2006) 6485 and references therein.

\bibitem{chen} CHEN F., MOHIDEEN U., KLIMCHITSKAYA G. L. and MOSTEPANENKO V.
M., \textit{Phys. Rev. Lett.} \textbf{88} (2002) 101801; \textit{Phys. Rev.}
A \textbf{66} (2002) 032113.

\bibitem{PFA} DERIAGIN B. V., \textit{Kolloid Z.} \textbf{69} (1934) 155;
DERIAGIN B. V., ABRIKOSOVA I. I. and LIFSHITZ E. M., \textit{Quart. Rev.} 
\textbf{10} (1968) 295.

\bibitem{EPL2003} GENET C., LAMBRECHT A., MAIA NETO, P. and REYNAUD S., 
\textit{Europhys. Lett.} \textbf{62} (2003) 484 .

\bibitem{Emig} EMIG T., HANKE A., GOLESTANIAN R. and KARDAR M., \textit{%
Phys. Rev.} A \textbf{67} (2003) 022114.

\bibitem{Genet2003} GENET C., LAMBRECHT A. and REYNAUD S., \textit{Phys. Rev.%
} A \textbf{67} (2003) 043811.

\bibitem{scattering} LAMBRECHT A., MAIA NETO P. A. and REYNAUD S., submitted
(2006).

\bibitem{PRA2005} MAIA NETO, P. A., LAMBRECHT A. and REYNAUD S.,  \textit{%
Phys. Rev.} A \textbf{72} (2005) 012115 .

\bibitem{PRL2006} RODRIGUES R. B., MAIA NETO P. A., LAMBRECHT A. and REYNAUD
S.,  \textit{Phys. Rev. Lett.} \textbf{96} (2006) 100402.

\bibitem{torsionbalances} GUNDLACH J.H., \textit{Meas. Sci. Technol.} 
\textbf{10} (1999) 454.

\bibitem{Parsegian} PARSEGIAN V. A. and WEISS G. H., \textit{J. Adhes.} 
\textbf{3} (1972) 259.

\bibitem{Barash} BARASH Y., \textit{Izv. Vyssh. Uchebn. Zaved.,
Radiofiz.} \textbf{12} (1978) 1637;.

\bibitem{vanEnk} van ENK S. J., \textit{Phys. Rev.} A 
\textbf{52} (1995) 2569.

\bibitem{Torres} TORRES-GUSM\'AN J. C. and MOCH\'AN W. L., \textit{%
J. Phys. A: Math Gen.} \textbf{39} (2006) 6791.

\bibitem{Iannuzzi} MUNDAY J. N., IANNUZZI D., BARASH Y. and CAPASSO F., 
\textit{Phys. Rev.} A \textbf{71} (2005) 042102.

\bibitem{foots} Whereas the method developed in Ref.~\cite{Emig} requires
the existence of a direction of translational symmetry (so as to allow for
the definition of field polarizations conserved in the scattering by the
surfaces), our approach allows for the calculation in a geometry with no
such symmetry, such as the rotated corrugated plates, because it explicitly
takes into account the coupling between different polarizations.

\end{thebibliography}
\end{document}